\documentclass[english]{article}
\usepackage[utf8]{inputenc}
\usepackage[T1]{fontenc}
\usepackage{babel}
\usepackage{amsmath}
\usepackage{graphicx}
\usepackage{fancyhdr}
\newcommand{\scidatalogo}{}
\newcommand{\overleaflogo}{}
\begin{document}

\title{\textbf{Substrate engineering for high quality emission of free and localized excitons from atomic monolayers in hybrid architectures}}

\author{Oliver Iff\textsuperscript{1+}, Yu-Ming He\textsuperscript{1{+}}, Nils Lundt\textsuperscript{1}, Sebastian Stoll\textsuperscript{1},\\
Vasilij Baumann\textsuperscript{1}, Sven H\"ofling\textsuperscript{1,2}, Christian Schneider\textsuperscript{1},}

\maketitle
\thispagestyle{fancy}

1.{Technische Physik and Wilhelm Conrad R\"ontgen Research Center for Complex Material Systems, Physikalisches Institut 
Universit\"at W\"urzburg, Am Hubland, D-97074 W\"urzburg, Germany \\
2. SUPA, School of Physics and Astronomy, University of St Andrews, St Andrews, KY16 9SS, United Kingdom \\
$^{+}$ These authors contributed equally\\
corresponding author:\\
Christian Schneider (christian.schneider@physik.uni-wuerzburg.de)

\begin{abstract}
Atomic monolayers represent a novel class of materials to study localized and free excitons in two dimensions and to engineer optoelectronic devices based on their significant optical response. Here, we investigate the role of the substrate on the photoluminescense response of MoSe$_2$ and WSe$_2$ monolayers exfoliated either on SiO$_2$ or epitaxially grown InGaP substrates. 
In the case of MoSe$_2$, we observe a significant qualitative modification of the emission spectrum, which is widely dominated by the trion resonance on InGaP substrates. However, the effects of inhomogeneous broadening of the emission features are strongly reduced. Even more strikingly, in sheets of WSe$_2$, we could routinely observe emission lines from localized excitons with linewidths down to the resolution limit of 70\,$\mu$ eV. This is in stark contrast to reference samples featuring WSe$_2$ monolayers on SiO$_2$ surfaces, where the emission spectra from localized defects are widely dominated by spectral diffusion and blinking behaviour. 
Our experiment outlines the enormous potential of III-V-monolayer hybrid architectures to obtain high quality emission signals from atomic monolayers, which are straight forward to integrate into nanophotonic and integrated optoelectronic devices.  
\end{abstract}

\section{Introduction}

Monolayers of transition metal dichalcogenides have moved into the focus of solid state spectroscopy, since these new materials feature a variety of unique optical properties. Monolayers composed of the transition metal Mo or W and the Chalcogene Se, S and Te crystallize in a honeycomb lattice which lacks an inversion center. This yields a characteristic bandstructure, where the direct bandgap transitions are located at the K and K' points of the hexagonal Brillouine zone. These points can be distinctly addressed by the polarization of an injection laser, which leads to novel spinor effects in these systems \cite{tmdreview, valley1, valley2, valley3, valley4, valley5}. In addition, so called valley excitons are formed, which feature an extraordinarily high binding energy exceeding 300 meV \cite{Ram2012}. This is a consequence of reduced dimensions, reduced dielectric screening and flat bands leading to a heavy exciton mass. In most of these materials, even up to ambient conditions, the absorption and luminescence spectrum is dominated by excitonic effects, rather than by direct interband transitions. While the general properties, such as the exciton frequency, and the trion binding energy are primarily determined by the monolayer itself, the surrounding environment still has a major influence on the optical properties. For instance, it has been shown that excitons in monolayers of MoS$_2$ sensibly react on absorbed molecules on the surface \cite{adsorb}, and energy shifts resulting from capping have been reported \cite{Sercombe2013}. Similarly, the choice of the substrate can have a significant effect on the luminescence properties of the free monolayer excitons, as well as the emission features from localized excitons which were recently identified as novel sources of single photon streams \cite{he2015single, koperski2015single, chakraborty2015voltage, srivastava2015optically, Tonndorf15}. 
Here, we study the excitonic properties of exfoliated monolayers of MoSe$_2$ and WSe$_2$ at cryogenic temperatures, which were transferred onto SiO$_2$/Si as well as InGaP/GaAs heterostructures. For the case of MoSe$_2$, we observe a strong reduction of the inhomogeneous broadening of the dominant trion feature as epitaxial substrates are utilized. In monolayers of WSe$_2$, we focus on the emission of localized excitons. This quantum dot like features are strongly broadened and disturbed by their environment on the insulating glass substrates. In stark contrast, the semiconducting InGaP/GaAs substrates facilitate dramatically reduced charge fluctuations, yielding stable and robust emitters of single photons on an epitaxial platform. 

\section{Sample structure and setup}
\label{sec:Sample structure and setup}

The investigated monolayers were produced by mechanical exfoliation from a MoSe$_2$ or a WSe$_2$ bulk crystal with scotch tape. After confirming their monolayer nature via their distinct photoluminescence and the color contrast in an optical microscope, they have been transferred onto the designated target substrate via a dry-stamp method \cite{transfer}.
Using this technique, flake sizes of around 30~$\mu$m~*~50~$\mu$m have been fabricated. Two different sample types have been implemented which are sketched in fig.~\ref{Fig1}\,(a). The monolayers were transferred onto substrates composed of a 90\,nm SiO$_2$ layer ontop of a Si substrate. The other substrate is made of a 250\,nm thick In$_{0.49}$Ga$_{0.51}$P layer which has been grown lattice-matched onto a semi-insulating GaAs by means of gas-source molecular beam epitaxy. In order to get an impression of the samples' surface quality we performed atomic force microscope (AFM) measurements, which can be seen in fig.~\ref{Fig1}\,(b). The  root-mean-squared roughness of both samples is of the same magnitude. Specifically, the SiO$_2$ surface characterized by a roughness of $\approx$ 0.15\,nm, while the InGaP surface features a comparable value of 0.29\,nm. 
Optical characterization was carried out in a standard microphotoluminescence setup ($\mu$PL). The samples were attached to the cold-finger of a liquid helium flow cryostat and the luminescence from the flake was collected by a 50x objective (NA=0.42) in a confocal microscope system. The structures were excited by a continuous-wave(cw) 532\,nm laser. Photoluminescence measurements were performed using a Princeton-Instrument SP2750i spectrometer equipped with a liquid nitrogen cooled charge coupled detector and a $1500~\frac{\textnormal{lines}}{\textnormal{mm}}$ grating ($\Delta E_{\textnormal{Res}}\approx70~$$\mu$eV) for the high resolution images or a $300~\frac{\textnormal{lines}}{\textnormal{mm}}$ grating for overview spectra. The PL could also be collected into a fiber coupled Hanbury Brown and Twiss setup (HBT) with a timing resolution of approximately 570~ps to measure the second order field correlation of the emission, after passing through a pair of band-pass filters (1\,nm bandwidth).

\section{Experimental Results and Discussion}
\label{sec:Experimental Results and Analysis}

First, we investigate the impact of the two aforementioned substrates on the emission characteristics of MoSe$_2$ monolayers. Fig.~\ref{Fig2} depicts a series of photoluminescence spectra of the Si/SiO$_2$-MoSe$_2$ structure which were recorded subsequently under nominally the same conditions and without blanking the laser. The spectra were taken in a time span of 10 minutes at a constant laser power of 50\,$\mu$W. We observe the common spectral signatures of MoSe$_2$ monolayers: At 1.657\,eV, the free exciton (X) is clearly visible. On the low energy side, the negatively charged trion (X-) emerges at 1.625\,meV, yielding a trion binding energy of 32\,meV. Noteworthily, during the series, the initial exciton intensity decreases and the trion intensity increases until both signals converge to a constant intensity ratio $I_{X-}/I_X \approx 2$ after roughly 5 minutes. This behaviour can be explained by a photo induced doping effect which introduces new free carriers into the system, enhancing the formation of trions \cite{Astakhov2002, trion} at the expense of free neutral excitons. On the contrary, on the GaAs/InGaP-MoSe$_2$ heterostructures, the free exciton is not visible while only the trion-attributed resonance can be clearly observed at an energy of 1.632\,eV at 50\,$\mu$W laser power. The spectral energy shift of approx $7$meV compared to the SiO$_2$-MoSe$_2$ stack occurs reproducibly in different flakes, and is most likely a consequence of the modified dielectric environment. Remarkably, the overall intensity of this trion resonance does not change with time, indicating that the system natively has access to a great amount of free carriers. We note, that both monolayers originate from the same bulk crystal and therefore we can rule  out inherent doping of the flake itself as a reason of this behavior.

In fig.~\ref{Fig3} we depict the result of a power series from both samples. The non-resonant excitation power was ramped from 50\,$\mu$W up to 5\,mW,  and we plot the trions' integrated intensity and the linewidth. With increasing power, the intensity of the observed resonances rises approximately linearly as shown in fig.~\ref{Fig3}(a). Fitting the data (red lines) to a straight line gives a slope of 0.96 for SiO$_2$ respectively 0.92 for InGaP, in good agreement with the expected slope of 1 for charged excitons. At higher output power (>\,1\,mW) the emissions start to show a saturation behaviour, independent
 from the used substrate caused by exciton annihilation \cite{annihil}. Another important parameter is the corresponding full width at half maximum (FWHM) of the studied signal, which is plotted in fig.~\ref{Fig3} (b). On the glass substrate the linewidth of the trion reaches a value around 13\,meV for small laser power. Increasing the power yields a progressive broadening of the emission line, reaching approximately 16\,meV at 6\,mW pump power. We assume, that this power induced broadening of the trion resonance is a consequence of local heating from the pump laser, but it could be also induced by additional charges which accumulate in the monolayer and at random positions at the heterointerface \cite{trion}. This charge puddling effect is known to occur on SiO$_2$ surfaces \cite{ChargeAcc}, which induce a randomly varying inhomogeneity on the photoluminescence response. Contrary to this, the linewidth on the InGaP sample is as small as 6.5\,meV, surpassing its SiO$_2$ counter part by a factor of 2. Even more remarkably, the linewidth stays nearly constant with increasing power and reaches just 7\,meV at 6\,mW laser output. This is due to a higher thermal conductivity of InGaP compared to SiO$_2$ \cite{thermal} leading to lower local heating at the laser spot. Overall, these results already outline the reduction of charge induced fluctuations in monolayer-InGaP devices, and illustrate the impact the right substrate can have on the excitonic properties of MoSe$_2$.

While monolayers of MoSe$_2$ are specificly suitable to study effects of free excitons and trions, the observation of single photon emission from localized excitons have set monolayers of WSe$_2$ in the focus of solid state quantum photonics. Figure 4 shows a typical photoluminescence spectrum from such a localized exciton in a WSe$_2$ monolayer ontop of a SiO$_2$/Si substrate, which was excited by a continuous-wave(CW) 532\,nm laser at an excitation power of 30\,$\mu$W and a nominal sample temperature of 4.2\,K. The photoluminescence spectrum consists of several sharp peaks with linewidths of ~2\,meV, centered at 1.52\,eV. Such a spectral feature, which is red-shifted 180\,meV from the WSe$_2$ free valley exciton(1.7eV), is comparable to previously reported localized emission signals in WSe$_2$ monolayers \cite{HeOptic}. Compared to the weak, broad PL spectrum from the localized exciton in the WSe$_2$ monolayer exfoliated on the SiO$_2$/Si substrate, several bright, spectral resolution limited (70\,$\mu$eV) PL peaks were observed from WSe$_2$ sheets transferred onto the InGaP/GaAs substrate (temperature of 4.5\,K). It is worth noting, that the extracted linewidth is more than 30 times smaller compared to the WSe$_2$-SiO$_2$ structure. Here, the PL excitation power for fig.~4 (b) is around 70\,nW, which is almost three order of magnitude smaller than the nominal 30\,$\mu$W for fig.~4 (a). Additionally, the right inset of fig.~4 (b) shows the cw-pumped autocorrelation histogram from the marked peak in fig.~4 (b). The emission is spectrally filtered by a pair of band-pass filters and then coupled into a fiber-based HBT-setup to measure the second order autocorrelation. The clear antibunching is observed around $\tau\approx0~$ns which reaches down well below $0.5$ and therefore proves the single photon emission. 

In order to account for the finite time resolution of our setup, we fit the measured data with a two sided exponential decay convolved with a Gaussian distribution $f_{Det}$:
\begin{eqnarray}
g^{(2)}_{source}(\tau)&=&1-((1-g^{(2)}(0))*e^{-|\frac{\tau}{\tau_C}|}) \\
g^{(2)}_{measured}(\tau)&=&(g^{(2)}_{source}\ast f_{Det})(\tau)
\label{g2fit}
\end{eqnarray}

%The convolved (blue) and deconvolved (red) fit functions are also plotted in the inset of fig.~\ref{Fig4} (b). 

Following this, we extract a deconvoluted $g^{(2)}(0)$-value of $g^{(2)}_{cw}=0.261\pm0.117$. \

To assess the influence of the spectral wandering on the macroscopic time scale on the emission features depicted in fig.~4 (a) and 4 (b), we record various spectra every second and combine them in the contour graph in fig.~5 (a) and 5 (b). In fig.5 (a), clear spectral wandering and jumps on the timescale of seconds are observed. Each frame is then fitted with a Lorentzian function and the statistics of the peak energies are plotted in fig.~5 (c). We find a direct contribution as large as $(957\pm58)$$\mu$eV of the long term spectral diffusion. This characteristic slow spectral jitter on such a large magnitude is commonly observed for self-assembled quantum emitters close to surfaces or interfaces which yield the capability of trapping and releasing charges. Thus, and in principle agreement with the studies presented in fig.~3 for the MoSe$_2$ case, we conclude that the spectral jumps are induced by carriers trapped via dangling bonds on the SiO$_2$ surface. Compared to the WSe$_2$ monolayer on SiO$_2$ substrate, no obvious spectral wandering is observed in fig.~5 (b), where the WSe$_2$ monolayer is transferred on the InGaP substrate. The corresponding statistics of the spectral wandering in fig.~5 (d) yields a value around $5.5$$\mu$eV, which is within the linewidth fitting uncertainty. The narrowing could be attributed to fewer charge fluctuation for semiconducting environment which allows transferring trapped charges, which leads to a suppression of the long scale spectral jitter.\\

Lastly, we performed a statistical study of the influence of the different substrate on the spectral linewidth of the localized excitons in the WSe$_2$ monolayers . A statistical histogram for 37 randomly localized emitters from 10 different monolayers on SiO$_2$/Si substrate is presented in fig.6 (a). The extracted linewidths randomly fluctuate between 147\,$\mu$eV to 3.3\,meV. 
Similarly, the statistical histogram for 259 randomly localized emitters from 10 different monolayers on InGaP/GaAs is depicted in fig.~6 (b). Here the extracted average linewidth is almost restricted by the spectrometer resolution (70\,$\mu$eV). Therefore, the resolution limited, jitter free PL strongly indicates that the InGaP substrate could greatly enhance the emission properties of the localized excitons in the WSe$_2$ monolayer.

\section{Summary}
\label{sec:Summary}

In conclusion, we have studied the influence of the substrate on the emission properties of monolayers of MoSe$_2$ and WSe$_2$ at cryogenic temperatures. On our reference SiO$_2$ substrate, the luminescence of the free exciton and trion in MoSe$_2$ is notably inhomogeneously broadened, and sensitive to power broadening. The investigated localized defects occuring in WSe$_2$ monolayers are subject to a long term spectral diffusion induced by a slowly varying charge environment. In stark contrast, InGaP substrates show a notable effect on the charge environment, which directly leads to a reduced broadening of the trionic emission in MoSe$_2$ and in many cases eliminates the slow spectral diffusion acting on localized emission centers in WSe$_2$. Together with the highly developed photonic processing technology of InGaP/GaAs structures, this makes WSe$_2$-InGaP heterostacks very interesting for novel nanophotonic and integrated monolayer based quantum photonic architectures. Furthermore, we have observed a significantly enhanced formation of free trions in MoSe$_2$ monolayers on InGaP, which make such a platform highly suitable to study interactions of monolayer excitations with electron gases, and can likely represent a new approach towards trion polaritons in a straight forward manner and without the necessity for electrostatic gating.  

\section*{Acknowledgments}

We acknowledge financial support by the State of Bavaria and the European Research Council (Project Unlimit-2D).

\newpage
\section*{Figures}

\begin{figure}[hbt]

\begin{center}
\includegraphics[width=1\textwidth]{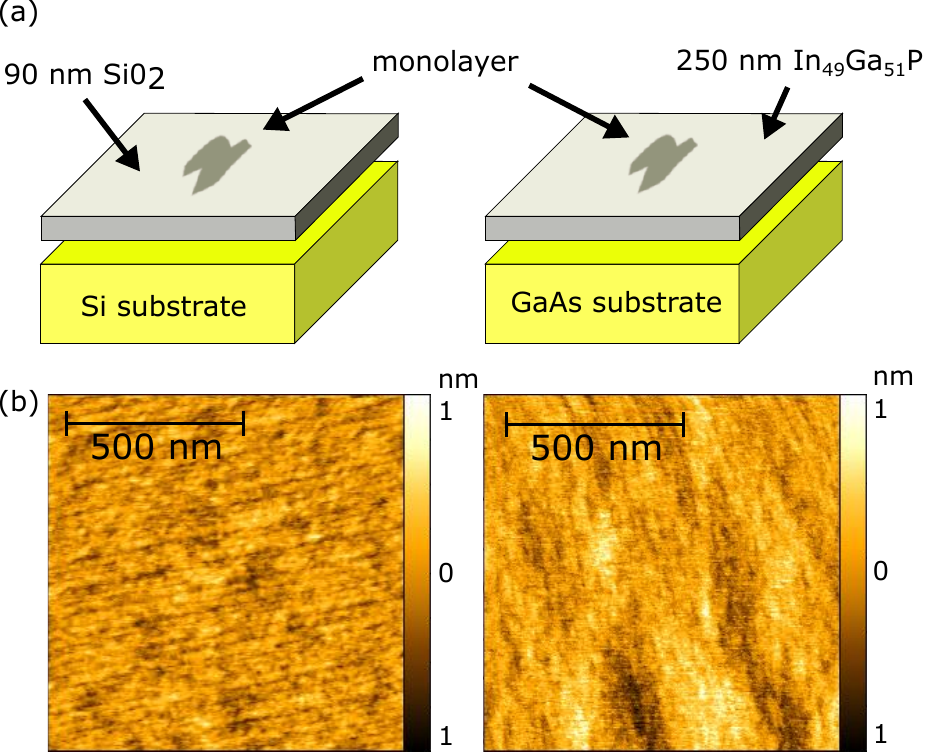}
\caption{(a)~Schematic drawing of the investigated heterostructures: 90\,nm SiO$_2$ on a Si substrate, and 250\,nm In$_{0.49}$Ga$_{0.51}$P lattice matched to a GaAs substrate. The monolayers were transfered onto each substrate, using the dry-stamp technique.  
 (b)~AFM measurements of the used samples. SiO$_2$ has a  root-mean-squared roughness of 0.15\,nm, while In$_{0.49}$Ga$_{0.51}$P has 0.29\,nm.}
\label{Fig1}
\end{center}
\end{figure}

\begin{figure}[hbt]
\begin{center}
\includegraphics[width=1\textwidth]{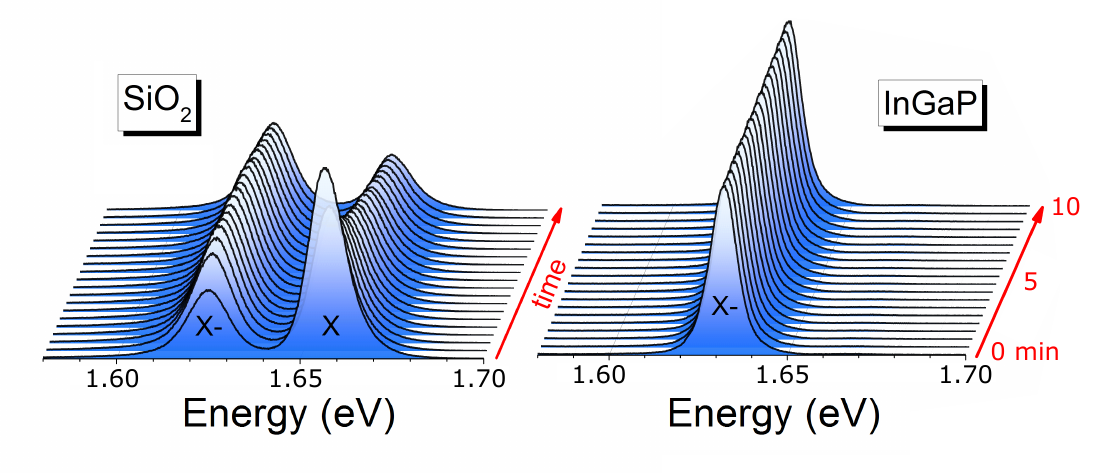}
\caption{Monolayer photoluminescence at 50\,$\mu$W, recorded over ten minutes. For MoSe$_2$ on SiO$_2$, the exciton intensity diminishes over time while the trion grows in intensity. For the MoSe$_2$-InGaP heterostructure , the trion dominates the spectrum by a large margin.}   
\label{Fig2}
\end{center}
\end{figure}

\begin{figure}[hbt]
\begin{center}
\includegraphics[width=1\textwidth]{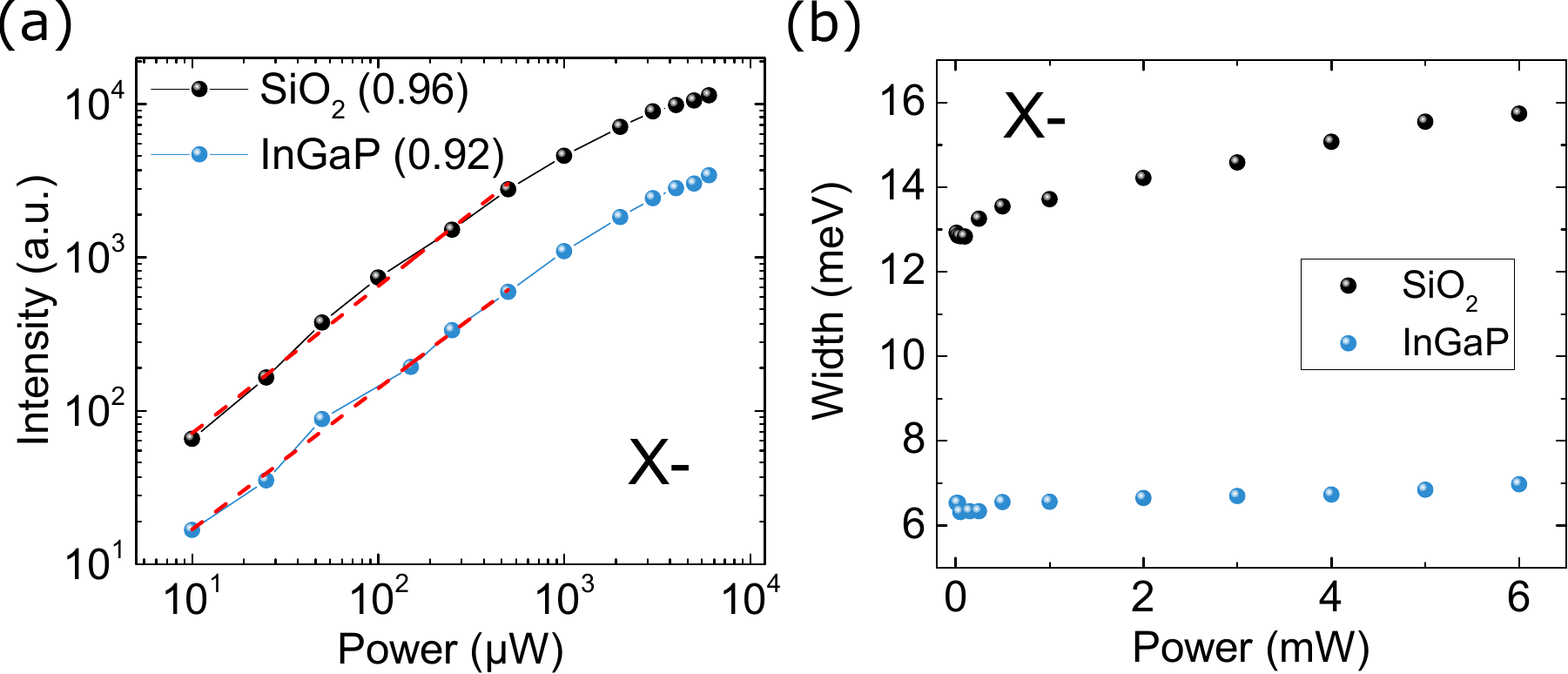}
\caption{(a)~Input-output characteristics of the trion intensity for MoSe$_2$ on SiO$_2$ and MoSe$_2$ on InGaP samples with an almost linear slope of 1. Dashed red lines are fitting curves. (b)~Corresponding FWHM of the trion. On SiO$_2$, it starts at 13\,meV, increasing at higher powers up to 16\,meV. On InGaP, the linewdith is 6.5\,meV, which stays almost constant with regard to laser output.}
\label{Fig3}
\end{center}
\end{figure}

\begin{figure}[hbt]
\begin{center}
\includegraphics[width=1\textwidth]{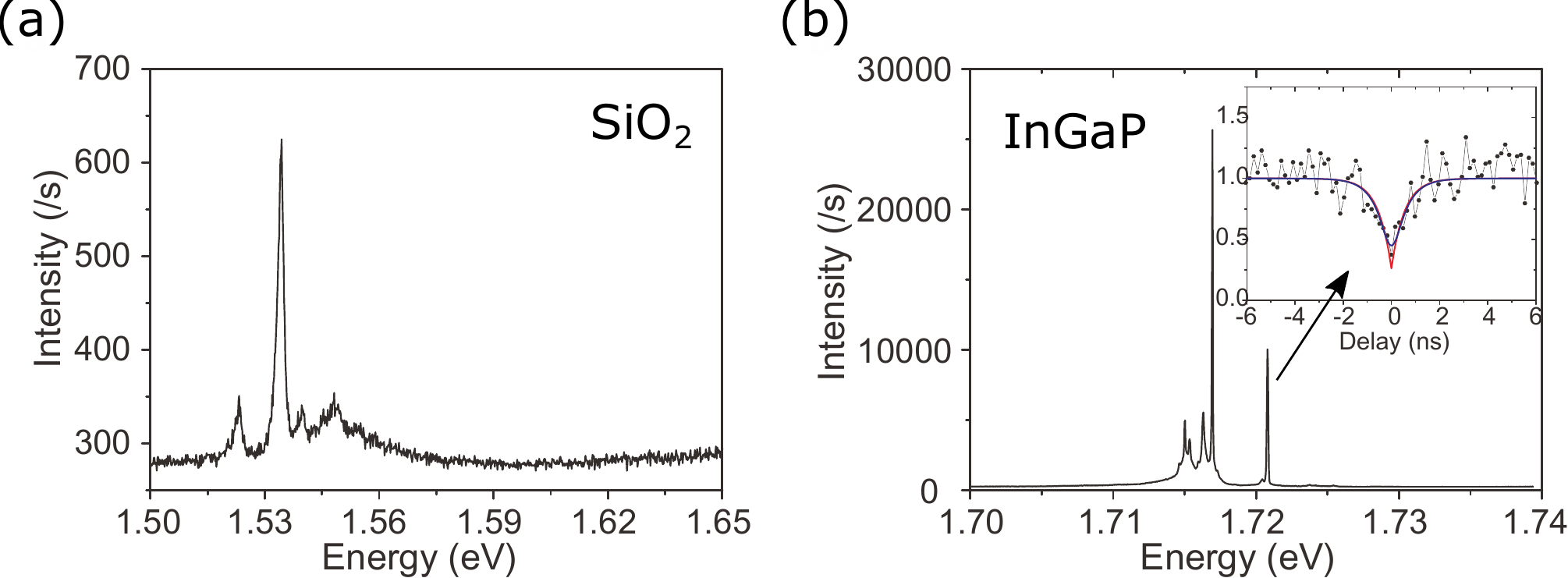}
\caption{(a) The typical PL spectrum of the localized exciton in the monolayer WSe$_2$ exfoliated onto a SiO$_2$/Si substrate, measured at a sample temperature of nominally 4.5\,K. (b) PL spectrum of the localized exciton in the monolayer WSe$_2$ with the InGaP/GaAs substrate under 4.5\,K. The peak energies range from 1.5\,eV to 1.73\,eV. The inset is the auto-correlation measurement of marked peak under a 70\,nW CW laser excitation at 532\,nm. The blue line in the inset is a fit with multiexcitonic model convolved with the response function. The red line is the deconvoluted curve, which shows $g^{(2)}$(0)=0.261$\pm$0.117.}
\label{Fig4}
\end{center}
\end{figure}

\begin{figure}[hbt]
\begin{center}
\includegraphics[width=1\textwidth]{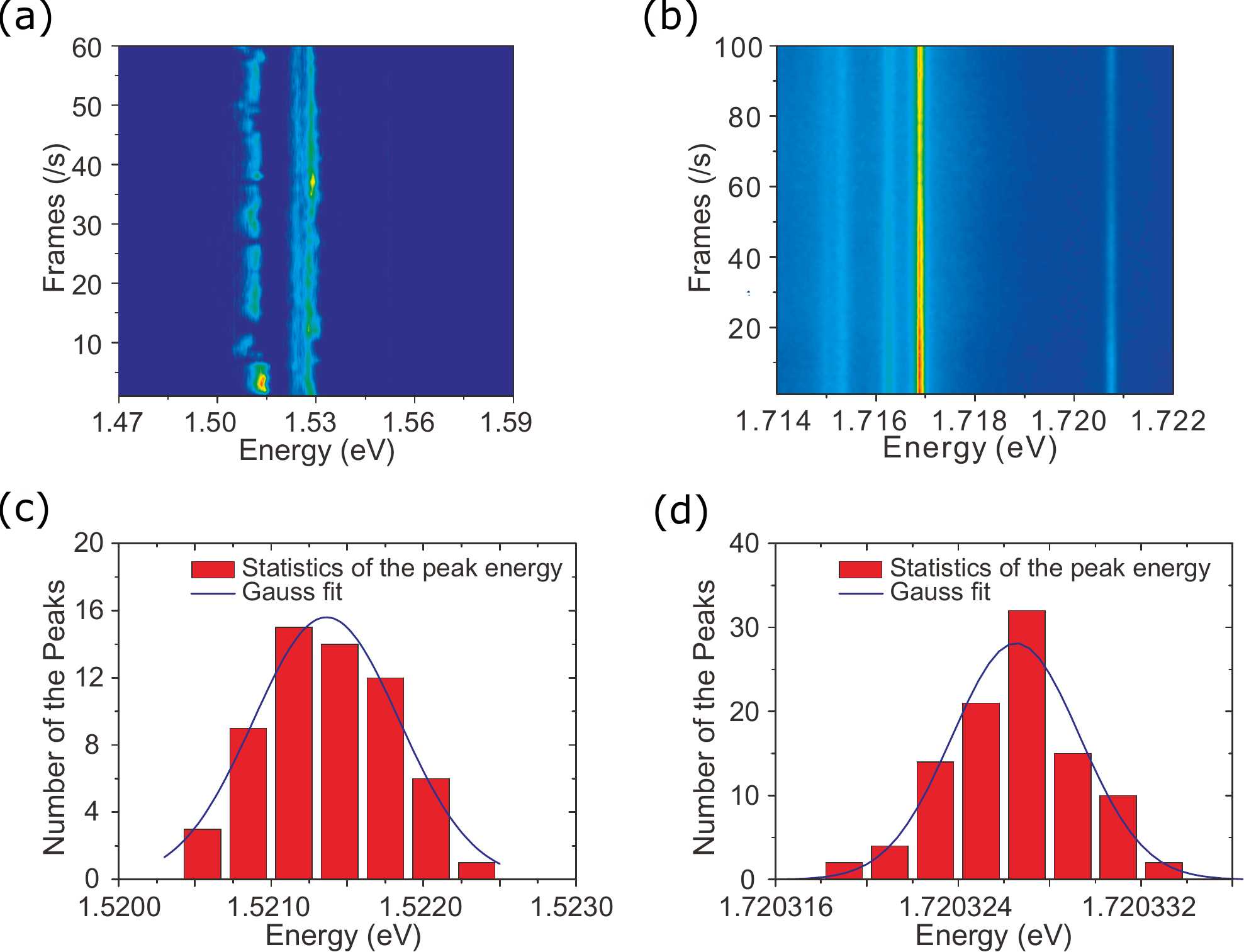}
\caption{(a) Spectral wandering of the localized exciton in layered WSe$_2$ on the SiO$_2$/Si substrate. (b) Emission time trace of the localized exciton in layered WSe$_2$ on the InGaP/GaAs substrate. Here, no obvious spectral wandering could be observed. (c) and (d) Statistics of the localized exciton peak A and peak B in Fig.~4(a) and Fig.4~(b) as a function of time. The extracted FWHM of the wandering are $(957\pm58)$$\mu$eV and $(5.583\pm0.582)$$\mu$eV respectively.}
\label{Fig5}
\end{center}
\end{figure}

\begin{figure}[hbt]
\begin{center}
\includegraphics[width=1\textwidth]{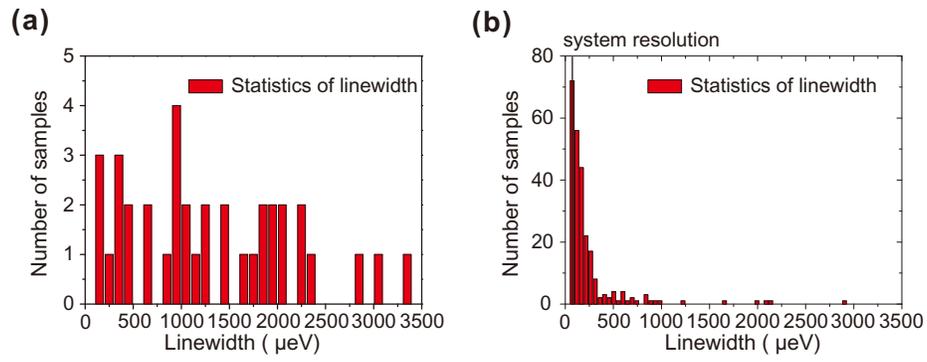}
\caption{(a) Statistic of the linewidth distribution for the 37 localized excitons in the WSe$_2$ monolayer on the SiO$_2$/Si substrate. The extracted minimum linewidth is $125$\,$\mu$eV. (b) Statistic of the linewidth distribution for the localized excitons in the WSe$_2$ monolayer on the InGaP/GaAs substrate. The average linewidth of $(74.8\pm12.2)$\,$\mu$eV of the 72 narrowest emission lines (first bin) is restricted by the resolution of our spectrometer ($70$\,$\mu$eV).}
\label{Fig6}
\end{center}
\end{figure}

\end{document}